\documentclass[superscriptaddress,twocolumn,showpacs,a4paper,
amssymb,amsmath,nobibnotes,aps,prl,
showkeys,
nofootinbib,notitlepage]{revtex4-1}
\usepackage{verbatim}
\usepackage[T1]{fontenc}
\usepackage[utf8]{inputenc}
\usepackage[american]{babel}
\usepackage{epsfig}
\usepackage{booktabs}
\usepackage{multirow}
\usepackage{dcolumn}
\usepackage{amsmath}
\usepackage{mathtools}
\usepackage{amsfonts}
\usepackage{amssymb}
\usepackage{ulem}
\usepackage{epstopdf}
\usepackage{bm}
\usepackage{siunitx}
\usepackage{braket}
\usepackage{enumitem}
\usepackage{soul}
\usepackage[table]{xcolor}
\usepackage{color}
\usepackage{transparent}
\usepackage{pifont}
\usepackage{enumitem}

\definecolor{navyblue}{rgb}{0.0, 0.0, 0.5}
\definecolor{royalblue}{rgb}{0.25, 0.41, 0.88}
\definecolor{cadmiumgreen}{rgb}{0.0, 0.42, 0.24}
\definecolor{blue-violet}{rgb}{0.54, 0.17, 0.89}
\definecolor{darkviolet}{rgb}{0.58, 0.0, 0.83}
\definecolor{orange(colorwheel)}{rgb}{1.0, 0.5, 0.0}

\usepackage{hyperref}
\hypersetup{
    colorlinks=true, 
    linkcolor=royalblue, 
    citecolor=magenta}

\usepackage{booktabs}
\usepackage{multirow}
\usepackage{dcolumn}
\usepackage{colortbl}

\begin{document}

\title{Interacting Dark Energy after DESI Baryon Acoustic Oscillation measurements}

\author{William Giar\`e}
\email{w.giare@sheffield.ac.uk}
\affiliation{School of Mathematics and Statistics, University of Sheffield, Hounsfield Road, Sheffield S3 7RH, United Kingdom}

\author{Miguel A. Sabogal}
\email{miguel.sabogal@ufrgs.br}
\affiliation{Instituto de F\'{i}sica, Universidade Federal do Rio Grande do Sul, 91501-970 Porto Alegre RS, Brazil}

\author{Rafael C. Nunes}
\email{rafadcnunes@gmail.com}
\affiliation{Instituto de F\'{i}sica, Universidade Federal do Rio Grande do Sul, 91501-970 Porto Alegre RS, Brazil}
\affiliation{Divis\~{a}o de Astrof\'{i}sica, Instituto Nacional de Pesquisas Espaciais, Avenida dos Astronautas 1758, S\~{a}o Jos\'{e} dos Campos, 12227-010, S\~{a}o Paulo, Brazil}

\author{Eleonora Di Valentino}
\email{e.divalentino@sheffield.ac.uk}
\affiliation{School of Mathematics and Statistics, University of Sheffield, Hounsfield Road, Sheffield S3 7RH, United Kingdom}

\begin{abstract}
\noindent
We investigate the implications of the Baryon Acoustic Oscillations measurement released by the Dark Energy Spectroscopic Instrument (DESI) for Interacting Dark Energy (IDE) models characterized by an energy-momentum flow from Dark Matter to Dark Energy. By combining Planck-2018 and DESI data, we observe a preference for interactions, leading to a non-vanishing interaction rate $\xi=-0.32^{+0.18}_{-0.14}$, which results in a present-day expansion rate $H_0=70.8^{+1.4}_{-1.7}$ km/s/Mpc, reducing the tension with the value provided by the SH0ES collaboration to less than $\sim 1.3 \sigma$. The preference for interactions remains robust when including measurements of the expansion rate $H(z)$ obtained from the relative ages of massive, early-time, and passively-evolving galaxies, as well as when considering distance moduli measurements from Type-Ia Supernovae sourced from the Pantheon-plus catalog using the SH0ES Cepheid host distances as calibrators. Overall, the IDE framework provides an equally good, or better, explanation of both high- and low-redshift \textit{background} observations compared to $\Lambda$CDM, while also yielding higher $H_0$ values that align more closely with the local distance ladder estimates. However, a limitation of the IDE model is that it predicts lower $\Omega_{m}$ and higher $\sigma_{8}$ values, which may not be fully consistent with large-scale structure data at the \textit{perturbation} level.
\end{abstract}

\keywords{}

\maketitle


The well-known discrepancy between the present-day expansion rate of the Universe ($H_0$) as measured by the SH0ES collaboration using local distance ladder measurements from Type Ia supernovae~\cite{Riess:2021jrx,Murakami:2023xuy,Breuval:2024lsv} ($H_0=73\pm1$ km/s/Mpc), and the value of the same parameter inferred by the Planck collaboration~\cite{Planck:2018vyg} from observations of temperature and polarization anisotropies in the Cosmic Microwave Background (CMB) radiation, assuming a $\Lambda$CDM cosmology ($H_0=67.4\pm0.5$ km/s/Mpc), has reached a statistical significance exceeding $5\sigma$. Barring any possible systematic origin of this discrepancy\footnote{This possibility appears increasingly unlikely given the extensive review of several potential sources of systematic error performed by the SH0ES collaboration~\cite{Riess:2021jrx,Riess:2024ohe,Brout:2023wol}.}, a fascinating possibility is that the Hubble tension might be pointing towards new physics beyond the standard $\Lambda$CDM model of cosmology.

Numerous theoretical attempts have been proposed to increase the value of $H_0$ inferred from CMB data and restore cosmic concordance~\cite{DiValentino:2021izs,Abdalla:2022yfr,Khalife:2023qbu,Schoneberg:2021qvd}. However, a compelling solution to the problem remains elusive. The primary challenge stems from the highly precisely determined angular scale of the acoustic peaks in the CMB spectra~\cite{Planck:2018vyg}. This scale sets the ratio between the sound horizon at recombination and the angular diameter distance to the last scattering surface. Increasing the value of $H_0$ without disrupting the acoustic scale requires either a reduction in the value of the sound horizon or a different post-recombination expansion history of the Universe able to compensate for a higher $H_0$ while preserving the angular diameter distance from the last scattering surface~\cite{Knox:2019rjx}.

Both of these possibilities face severe constraints. Reducing the value of the sound horizon requires new physics acting at very high redshifts, typically just prior to recombination. Even ignoring the common fine-tuned problems surrounding early-time solutions, they remain severely constrained by high redshift observations, most notably by CMB data~\cite{Vagnozzi:2023nrq}. Conversely, late-time solutions require new physics altering cosmic distances to compensate for the higher values of $H_0$ while preserving the angular diameter distance from the CMB. In turn, cosmic distances are precisely measured by Baryon Acoustic Oscillations (BAO) and Type-Ia supernovae (SN) data that so far have not provided any evidence for deviations from a late-time $\Lambda$CDM cosmology, significantly reducing the room allowed for new physics at low redshift~\cite{Efstathiou:2021ocp,Krishnan:2021dyb,Keeley:2022ojz,Gariazzo:2024sil}.

\begin{table*}[htpb!]
\begin{center}
\renewcommand{\arraystretch}{1.5}
\resizebox{\textwidth}{!}{
\begin{tabular}{l c c c c c c c c c c c c c c c }
\hline
\textbf{Parameter} & \textbf{ Planck-2018+DESI } & \textbf{ Planck-2018+DESI+CC } & \textbf{ Planck-2018+DESI+SN } & \textbf{ Planck-2018+DESI+SN+CC } \\ 
\hline\hline

$ \Omega_\mathrm{b} h^2  $ & $  0.02243\pm 0.00014\, ( 0.02243^{+0.00028}_{-0.00026} ) $ & $  0.02243\pm 0.00014\, ( 0.02243^{+0.00027}_{-0.00027} ) $ & $  0.02254\pm 0.00013\, ( 0.02254^{+0.00026}_{-0.00027} ) $ & $  0.02255\pm 0.00014\, ( 0.02255^{+0.00027}_{-0.00027} ) $ \\ 
$ \Omega_\mathrm{c} h^2  $ & $  0.079^{+0.025}_{-0.016}\, ( 0.079^{+0.037}_{-0.042} ) $ & $  0.080^{+0.025}_{-0.016}\, ( 0.080^{+0.037}_{-0.042} ) $ & $  0.0962^{+0.0085}_{-0.0074}\, ( 0.096^{+0.015}_{-0.015} ) $ & $  0.0966^{+0.0084}_{-0.0075}\, ( 0.097^{+0.015}_{-0.016} ) $ \\ 
$ 100\theta_\mathrm{s}  $ & $  1.04198\pm 0.00029\, ( 1.04198^{+0.00056}_{-0.00056} ) $ & $  1.04197\pm 0.00028\, ( 1.04197^{+0.00054}_{-0.00056} ) $ & $  1.04211\pm 0.00028\, ( 1.04211^{+0.00055}_{-0.00057} ) $ & $  1.04211\pm 0.00028\, ( 1.04211^{+0.00054}_{-0.00054} ) $ \\ 
$ \tau_\mathrm{reio}  $ & $  0.0555\pm 0.0074\, ( 0.055^{+0.015}_{-0.014} ) $ & $  0.0554^{+0.0069}_{-0.0078}\, ( 0.055^{+0.016}_{-0.014} ) $ & $  0.0592^{+0.0069}_{-0.0079}\, ( 0.059^{+0.016}_{-0.014} ) $ & $  0.0590\pm 0.0077\, ( 0.059^{+0.016}_{-0.015} ) $ \\ 
$ n_\mathrm{s}  $ & $  0.9672\pm 0.0037\, ( 0.9672^{+0.0073}_{-0.0072} ) $ & $  0.9673\pm 0.0037\, ( 0.9673^{+0.0074}_{-0.0073} ) $ & $  0.9696\pm 0.0038\, ( 0.9696^{+0.0075}_{-0.0073} ) $ & $  0.9693\pm 0.0038\, ( 0.9693^{+0.0073}_{-0.0076} ) $ \\ 
$ \log(10^{10} A_\mathrm{s})  $ & $  3.045\pm 0.014\, ( 3.045^{+0.029}_{-0.028} ) $ & $  3.045^{+0.014}_{-0.015}\, ( 3.045^{+0.030}_{-0.028} ) $ & $  3.051\pm 0.015\, ( 3.051^{+0.031}_{-0.028} ) $ & $  3.050^{+0.014}_{-0.016}\, ( 3.050^{+0.031}_{-0.029} ) $ \\ 
$ \xi $ & $  -0.32^{+0.18}_{-0.14}\, ( -0.32^{+0.30}_{-0.29} ) $ & $  -0.32^{+0.18}_{-0.14}\, ( -0.32^{+0.30}_{-0.28} ) $ & $  -0.186\pm 0.068\, ( -0.19^{+0.13}_{-0.14} ) $ & $  -0.183\pm 0.069\, ( -0.18^{+0.13}_{-0.14} ) $ \\ 
$ H_0  $ [km/s/Mpc] & $  70.8^{+1.4}_{-1.7}\, ( 70.8^{+2.8}_{-2.7} ) $ & $  70.7^{+1.4}_{-1.7}\, ( 70.7^{+2.9}_{-2.7} ) $ & $  69.87\pm 0.60\, ( 69.9^{+1.2}_{-1.2} ) $ & $  69.83\pm 0.59\, ( 69.8^{+1.2}_{-1.1} ) $ \\ 
$ \Omega_\mathrm{m}  $ & $  0.206^{+0.056}_{-0.044}\, ( 0.206^{+0.090}_{-0.096} ) $ & $  0.208^{+0.057}_{-0.043}\, ( 0.208^{+0.090}_{-0.097} ) $ & $  0.245\pm 0.020\, ( 0.245^{+0.037}_{-0.039} ) $ & $  0.246\pm 0.020\, ( 0.246^{+0.038}_{-0.040} ) $ \\ 
$ \sigma_8  $ & $  1.23^{+0.14}_{-0.36}\, ( 1.23^{+0.74}_{-0.52} ) $ & $  1.220^{+0.082}_{-0.36}\, ( 1.22^{+0.70}_{-0.45} ) $ & $  0.974^{+0.059}_{-0.088}\, ( 0.97^{+0.15}_{-0.14} ) $ & $  0.971^{+0.060}_{-0.086}\, ( 0.97^{+0.15}_{-0.14} ) $ \\ 
$ r_\mathrm{drag}  $ [Mpc] & $  147.28\pm 0.23\, ( 147.28^{+0.45}_{-0.45} ) $ & $  147.28\pm 0.24\, ( 147.28^{+0.46}_{-0.46} ) $ & $  147.42\pm 0.23\, ( 147.42^{+0.44}_{-0.46} ) $ & $  147.40\pm 0.23\, ( 147.40^{+0.44}_{-0.45} ) $ \\ 
\hline
$\Delta \chi^2$ & $-1.02$ & $-2.00$ & $-2.27$ & $-2.21$\\
$\ln \mathcal B_{ij}$ & $-0.10$ & $-0.47$ & $-0.32$ & $-0.01$\\
\hline \hline
\end{tabular} }
\end{center}
\caption{Constraints at $68\%$ ($95\%$) CL on the parameters of the IDE model. For all datasets, we provide $\Delta \chi^2 = \chi^2_{\rm{IDE}} - \chi^2_{\Lambda\rm{CDM}}$ as well as the Bayes factors $\ln \mathcal{B}_{ij} = \ln \mathcal{Z}_{\Lambda\text{CDM}} - \ln \mathcal{Z}_{\text{IDE}}$ calculated as the difference between the evidence for $\Lambda$CDM and IDE model. Negative values of $\Delta \chi^2$ and $\ln \mathcal{B}_{ij}$ indicate a better fit and a preference for the IDE model over the $\Lambda$CDM, respectively.}
\label{tab.results}
\end{table*}

Interestingly, recent BAO measurements released by the Dark Energy Spectroscopic Instrument~\cite{DESI:2024uvr,DESI:2024lzq,desicollaboration2024desi} (DESI) appear to point towards new physics in the dark energy sector of the cosmological model~\cite{desicollaboration2024desi}. Following the intrinsic interest sparked by these new observations~\cite{Wang:2024hks,Cortes:2024lgw,Colgain:2024xqj,Yin:2024hba}, in this \textit{letter}, we examine their implications for cosmological models known as Interacting Dark Energy (IDE) where a non-gravitational interaction between Dark Matter (DM) and Dark Energy (DE) is postulated. Over the years, these models have been extensively explored as a potential avenue for resolving cosmological tensions~\cite{Pourtsidou:2016ico,DiValentino:2017iww,Kumar:2017dnp,vonMarttens:2019ixw,Lucca:2020zjb,Zhai:2023yny,Bernui:2023byc,Hoerning:2023hks,Giare:2024ytc}. Despite high-redshift data supporting IDE as solutions to the Hubble tension~\cite{Zhai:2023yny}, the situation remains somewhat unclear when examining low-redshift observations, as different probes yield somewhat discordant conclusions~\cite{Giare:2024ytc,Nunes:2022bhn,Yang:2022csz}. In this \textit{letter}, we demonstrate that the new DESI data give a preference for interactions exceeding the 95\% confidence level (CL) and that high- and low-redshift background observations can be equally or better explained in IDE than in $\Lambda$CDM, while yielding higher values of $H_0$ compatible with SH0ES.

\bigskip

We consider a homogeneous and isotropic Universe and introduce an energy-momentum flow in the dark sector of the model by modifying the energy-momentum equation as
\begin{equation}
\label{DE_DM_1}
\nabla_{\mu}T_{i}^{\mu \nu }=Q_{i}^{\nu}, \quad \sum\limits_{i}{Q_{i}^{\mu }}=0~.
\end{equation}
The degree of interaction is quantified by the four-vector
\begin{equation}
Q_{i}^{\mu}=(Q_{i}+\delta Q_{i})u^{\mu}+a^{-1}(0,\partial^{\mu}f_{i}),
\end{equation}
where $u^{\mu}$ represents the velocity four-vector, $Q_i$ is the background energy transfer, and the index $i$ runs over DM and DE. We adopt a widely recognized interaction kernel $Q = \mathcal{H} \xi \rho_{\rm DE}$~\cite{Gavela:2010tm,DiValentino:2019jae,Zhai:2023yny,silva2024nonlinear,DiValentino:2019ffd} where $\mathcal{H}$ is the (conformal) Hubble parameter, $\rho_{\rm DE}$ is the DE energy-density and $\xi$ dictates both the amount and the direction of energy-momentum flow. We require $\xi < 0$, forcing the energy-momentum transfer from DM to DE. Additionally, we fix the DE equation of state to $w\simeq-1$, resembling an interacting vacuum scenario.~\footnote{We regularize early-time super-horizon instabilities in the dynamics of cosmological perturbations~\cite{Valiviita:2008iv,Gavela:2010tm} by setting $w=-1+\epsilon$ and taking $\epsilon~\simeq 0.0001 \to0$.}

\bigskip
We implement the theoretical model in a modified version of the Boltzmann solver code \texttt{CLASS}~\cite{Blas:2011rf} and use the publicly available sampler \texttt{COBAYA}~\cite{Torrado:2020dgo} to perform Markov Chain Monte Carlo (MCMC) analyses. We assume flat priors on the set of sampled cosmological parameters \{$\Omega_b h^2$, $\Omega_c h^2$, $\tau_{\rm reio}$, $\theta_{\mathrm{s}}$, $\log(10^{10} A_{\mathrm{s}})$, $n_{\mathrm{s}}$, $\xi$\}. Our baseline datasets include the \textbf{Planck-2018} CMB temperature polarization and lensing likelihoods~\cite{Planck:2019nip,Planck:2018vyg,Planck:2018nkj} and the \textbf{DESI} BAO measurements obtained from observations of galaxies $\&$ quasars~\cite{DESI:2024uvr}, and Lyman-$\alpha$~\cite{DESI:2024lzq} tracers summarized in Tab.~I of Ref.~\cite{desicollaboration2024desi}. These latter are characterized in terms of measurements of the transverse comoving distance ($D_{\mathrm{M}}/r_{\mathrm{d}}$), the Hubble horizon ($D_{\mathrm{H}}/r_{\mathrm{d}}$), and the angle-averaged distance ($D_{\mathrm{V}}/r_{\mathrm{d}}$), normalized to the (comoving) sound horizon at the drag epoch, $r_{\mathrm{d}}$. We account for the correlation between measurements of $D_{\mathrm{M}}/r_{\mathrm{d}}$ and $D_{\mathrm{H}}/r_{\mathrm{d}}$. In addition to CMB and BAO data, we also consider distance moduli measurements from Type Ia \textbf{SN} gathered from the Pantheon-plus sample~\cite{Brout:2022vxf}. For this latter we use the SH0ES Cepheid host distances as calibrators~\cite{Riess:2021jrx}. Finally, we include measurements of the expansion rate $H(z)$ derived from the relative ages of massive, early-time, passively-evolving galaxies, known as cosmic chronometers (\textbf{CC})~\cite{Jimenez:2001gg}. We conservatively use only 15 CC measurements in the redshift range $0.179<z<1.965$~\cite{Moresco:2012by,Moresco:2015cya,Moresco:2016mzx}, accounting for all non-diagonal terms in the covariance matrix and systematic contributions.


\begin{figure}[htpb!]
    \centering
    \includegraphics[width=\columnwidth]{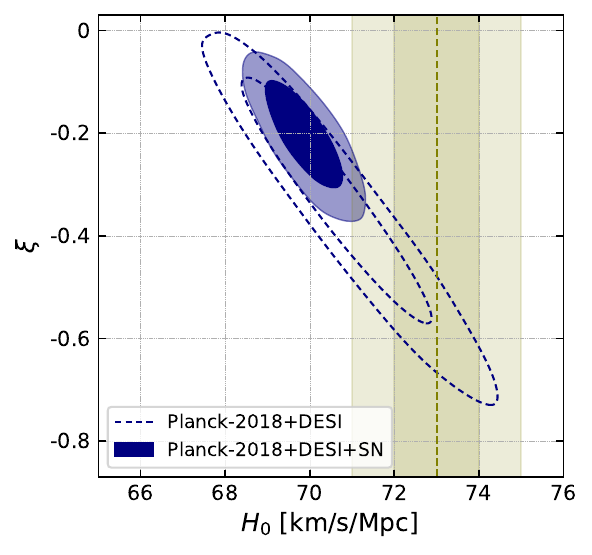} \,\,\,
    \caption{2D contours at $68\%$ and $95\%$ CL for the coupling parameter $\xi$ and the Hubble parameter $H_0$, as inferred by the different combinations of Planck-2018, DESI, and SN data listed in the legend. The olive-green band represents the value of $H_0$ measured by the SH0ES collaboration.}
    \label{fig:H0}
\end{figure}

\begin{figure}[htpb!]
    \centering
    \includegraphics[width=\columnwidth]{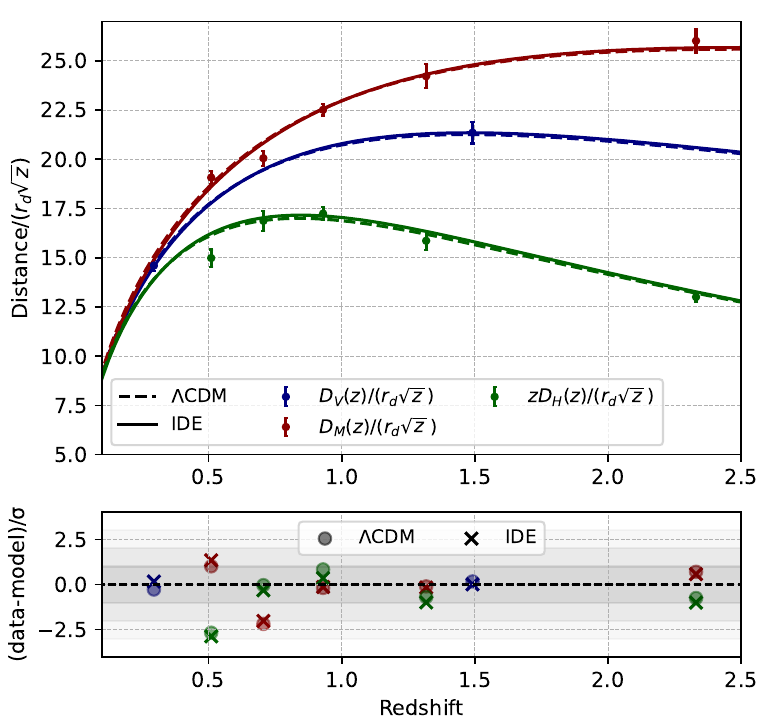} \,\,\,
    \caption{\textit{Upper panel}: Best-fit predictions for (rescaled) distance-redshift relations for IDE (solid curves) and $\Lambda$CDM (dashed curves) obtained from the analysis of Planck-2018+DESI data. These predictions are presented for the three different types of distances probed by BAO measurements, each indicated by the colors reported in the legend. The error bars represent $\pm 1 \sigma$ uncertainties. \textit{Lower panel}: Difference between the model prediction and data-point for each BAO measurement, normalized by the observational uncertainties. The IDE predictions are represented by 'x'-shaped points while the $\Lambda$CDM predictions are represented by 'o'-shaped points.}
    \label{fig:BAO}
\end{figure}

\bigskip

We summarize the constraints on cosmological parameters at 68\% and 95\% CL in Tab.~\ref{tab.results}. The most important results read as follows:
\begin{itemize}[leftmargin=*]

\item The joint Planck+DESI analysis yields a preference for a non-vanishing \textbf{$\xi = -0.32^{+0.18}_{-0.14}$}, exceeding the 95\% CL. Additionally, it provides a value $H_0 = 70.8^{+1.4}_{-1.7}$ km/s/Mpc, in agreement with local distance ladder estimates within $\sim 1.3 \sigma$. Therefore, focusing on Planck-2018 and DESI-BAO altogether, IDE can fully resolve the Hubble tension, see also Fig.~\ref{fig:H0}. Adding CC does not change this result, see also Tab.~\ref{tab.results}. That said, as shown in Table~\ref{tab.results}, the IDE model predicts lower $\Omega_{m}$ and higher $\sigma_{8}$ values, which suggest potential challenges in providing a consistent description of large-scale structure data at the perturbation level. We discuss this aspect in more detail at the end of this \textit{Letter}, where we analyze the impact of Redshift Space Distortion (RSD) measurements and highlight some limitations of the IDE framework.

\item Combining Planck-2018+DESI+SN, we still find a preference for $\xi \neq 0$ at more than 95\% CL, consistently yielding values of $H_0$ higher than in the standard cosmological model, see also Tab.~\ref{tab.results} and Fig.~\ref{fig:H0}. Again, this conclusion does not change considering CC. A non-vanishing energy-momentum flow from DM to DE is potentially supported by \textit{all} the main high and low redshift background measurements analyzed in this work.

\item For all the datasets listed in Tab.~\ref{tab.results}, we compare the best-fit $\chi^2_{\rm{IDE}}$ obtained within IDE to the best-fit $\chi^2_{\Lambda\rm{CDM}}$ obtained for $\Lambda$CDM. We find that $\Delta \chi^2 = \chi^2_{\rm{IDE}} - \chi^2_{\Lambda\rm{CDM}}$ is always negative. This means that IDE can fit data better than $\Lambda$CDM for \textit{all} the different combinations of data while simultaneously yielding higher values of $H_0$.

\item To account for the fact that IDE has one more free parameter than $\Lambda$CDM, we perform a model comparison and calculate the Bayes factors $\ln \mathcal B_{ij}$ normalized (for each dataset) to a baseline $\Lambda$CDM scenario in such a way that a negative $\ln \mathcal B_{ij}$ indicate a preference for IDE over $\Lambda$CDM, and vice versa.\footnote{To do so, we employ the \texttt{MCEvidence} package, which is publicly available~\cite{Heavens:2017hkr,Heavens:2017afc} and can be accessed at the following link: \url{https://github.com/yabebalFantaye/MCEvidence}.} Despite a trend towards $\ln \mathcal B_{ij} < 0$, the evidence is always inconclusive. Therefore, we conclude that IDE and $\Lambda$CDM can be deemed equally plausible to fit current observations.

\end{itemize}

\begin{table}[tpb!]
\begin{center}
\renewcommand{\arraystretch}{1.5}
\resizebox{0.97 \columnwidth}{!}{
\begin{tabular}{l c c c c}
\hline
 \textbf{Distance} & \textbf{Redshift} & \textbf{DESI} & \textbf{IDE ($\#\sigma$)} & \textbf{$\Lambda$CDM ($\#\sigma$) } \\ 
 \hline
 
&  & & \\

$D_V(z)/(r_d \sqrt{z})$ & $0.295$ & $14.60 \pm 0.28$ & $+0.16\, \sigma$ & $-0.28\, \sigma$ \\

& $1.491$ & $21.35 \pm 0.55$ & $+0.01\,\sigma$ & $+0.17\,\sigma$\\

&  & & \\

$D_M(z)/(r_d \sqrt{z})$ & $0.510$ & $19.07 \pm 0.35$ & $+1.34 \, \sigma$ & $+1.01 \, \sigma$\\
& $0.706$ & $20.05 \pm 0.38$ & $-2.02\, \sigma$ & $-2.18\, \sigma$\\
& $0.930$ & $22.51 \pm 0.29$ & $-0.14\, \sigma$ & $-0.19\, \sigma$\\
& $1.317$ & $24.22 \pm 0.60$ & $-0.17\, \sigma$ & $-0.11\, \sigma$\\
& $2.330$ & $26.01 \pm 0.61$ & $+0.57\, \sigma$ & $+0.70\, \sigma$\\

&  & & \\

$z D_H(z) / (r_d\sqrt{z})$ & $0.510$ & $14.98 \pm 0.43$ & $-2.87\sigma$ & $-2.67\sigma$\\
& $0.706$ & $16.87 \pm 0.50$ & $-0.32\, \sigma$ & $-0.03\, \sigma$\\
& $0.930$ & $17.24 \pm 0.34$ & $+0.35\, \sigma$ & $+0.84\, \sigma$\\
& $1.317$ & $15.86 \pm 0.48$ & $-0.99\, \sigma$ & $-0.69\, \sigma$\\
& $2.330$ & $ 13.00 \pm 0.25$ & $-1.00\, \sigma$ & $-0.75\, \sigma$\\

\hline \hline
\end{tabular}
}
\end{center}
\caption{The DESI results (and their 1$\sigma$ errors) are presented for three distinct types of distances investigated by BAO measurements. For each data point, we indicate the consistency between the best-fit predictions of the IDE and $\Lambda$CDM models and the observed data, expressed in units of observational uncertainties ($\# \sigma$).}
\label{tab.BAOs}
\end{table}

Taking the results obtained by combining Planck-2018 and DESI BAO distance measurements at face value, there is solid ground to conclude that they lend weight to the possibility of non-vanishing energy-momentum transfer from DM to DE, resulting in the pronounced negative correlation between $\xi$ and $H_0$ depicted in Fig.~\ref{fig:H0}. This preference can be better understood by referring to Ref.~\cite{desicollaboration2024desi}. In this work, the DESI collaboration argued that considering a dynamical $w_0w_a$CDM parameterization, Planck-2018+DESI data produce a strong preference for a (dynamical) DE equation of state showing a phantom behavior in the past. In IDE, upon straightforward manipulation of the continuity equation, one can rearrange the effective equation of state parameters to be $w_{\text{eff}} = -1 + \xi/3$. Consequently, the preference for a late-time phantom-like behavior is recast into an indication $\xi < 0$ here exceeding the 95\% CL. Focusing on Planck+DESI(+CC), this preference can fully resolve the Hubble tension. However, the significant fraction of energy-momentum transferred from DM to DE naturally implies lower values of the matter density parameter ($\Omega_m$), and predicts a larger value of the variance in the mass distribution smoothed over a spherical volume of radius $R = 8 h^{-1}$Mpc ($\sigma_{8}$) compared to the standard cosmological model. When SN data are included in the analysis, we observe a tendency towards higher values of $\Omega_m$ that reduces the value inferred for $H_0$. Having said that, despite a larger $\Omega_m$, $H_0$ remains in better agreement with SH0ES, reducing the Hubble tension to $2.6 \sigma$. Therefore, while within the $\Lambda$CDM model, forcing $H_0$ to move towards the SH0ES value leaves little room for agreement with local distance ladder estimates, IDE could offer a more flexible theoretical framework.\footnote{We note that this essentially means the model can fit both the LSS data and SH0ES measurements simultaneously, without resulting in a poor fit. If one attempted the same with $\Lambda$CDM, it would also reduce the Hubble tension, but at the cost of significantly increasing the $\chi^2$ (by $\sim 25$) due to the discrepancy between the cosmological $H_0$ inference and the SH0ES measurement.}

To better understand the role played by DESI data, we compare the theoretical distance predictions for IDE and $\Lambda$CDM against the observed cosmic distances. In particular, in Fig.~\ref{fig:BAO}, we compare the Planck-2018+DESI best-fit predictions for the three different types of (rescaled) distances probed by BAO measurements. In the bottom panel of the same figure, we show the distance between the observed DESI data points and the best-fit predictions obtained for $\Lambda$CDM (‘o’-shaped points) and IDE (‘x’-shaped points) in units of observational uncertainty $\sigma$. The same difference between the model predictions and DESI data is summarized in Tab.~\ref{tab.BAOs}. Comparing the best-fit predictions for IDE and $\Lambda$CDM, some important conclusions can be reached. Foremost, we observe that the DESI data point showing the largest disagreement with IDE (at a level of approximately $2.9\sigma$) is the measurement of $z D_H(z) / (r_d \sqrt{z})$ at $z=0.510$. The only other DESI BAO measurement displaying a significant deviation from the IDE best-fit curve (at approximately $2\sigma$) is $D_M(z) / (r_d \sqrt{z})$ at $z=0.706$. However, these two data points also exhibit significant disagreement with $\Lambda$CDM (at approximately $2.7\sigma$ and $2.2\sigma$, respectively). Therefore, IDE explains $D_M(z) / (r_d \sqrt{z})$ at $z=0.706$ more successfully than $\Lambda$CDM. Overall, apart from these two distance measurements at $z=0.51$ and $z=0.706$ (which -- repetita iuvant  --  are also at odds with $\Lambda$CDM), there are no other BAO measurements in tension with the IDE best-fit predictions. Last but not least, we stress that a large part of the improvement in the total $\Delta \chi^2 = -1.02$ over $\Lambda$CDM, as shown in Table~\ref{tab.results} for Planck-2018+DESI, comes specifically from the DESI BAO measurement ($\Delta \chi^2_{\rm{DESI}} = -0.73$).

\begin{figure}[tpb!]
    \centering
    \includegraphics[width=\columnwidth]{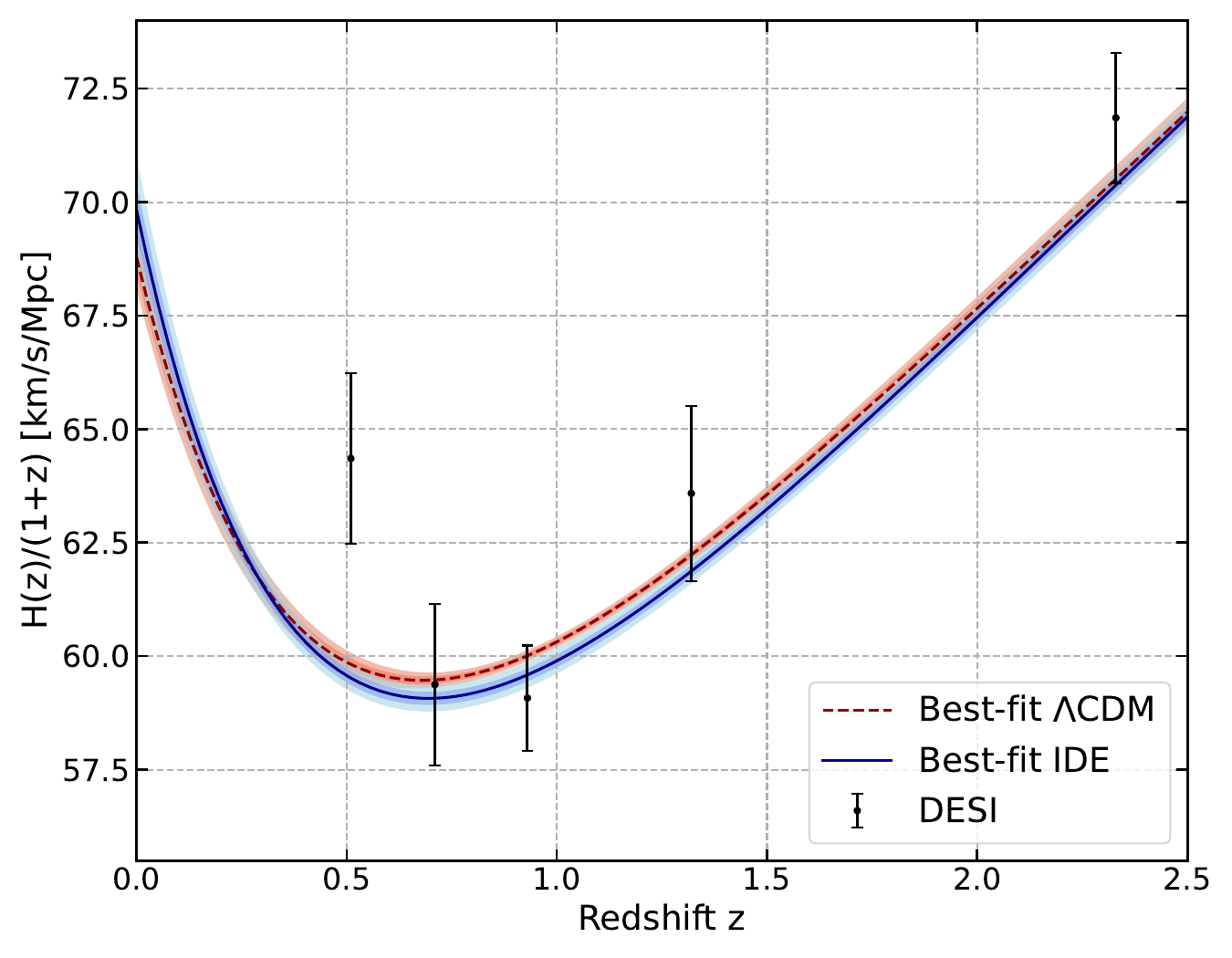} 
    \caption{Statistical reconstruction of the (rescaled) expansion rate of the universe $H(z)/(1+z)$ at 1$\sigma$ and 2$\sigma$ confidence levels for $\Lambda$CDM and IDE models through the joint analysis of Planck-2018+DESI+SN+CC, compared to DESI measurements.}
    \label{fig:H(z)}
\end{figure}

In Fig.~\ref{fig:H(z)}, we present the theoretical predictions for $H(z)$ as obtained by simultaneously analyzing Planck-2018, DESI, CC, and SN. We display the best-fit predictions (along with their 1 and 2$\sigma$ uncertainties) for IDE and $\Lambda$CDM, comparing them against the data points released by DESI. As evident from the figure, even accounting for all datasets together, the DESI datapoint at $z=0.51$ remains essentially unexplained in both models. Conversely, IDE can fit $H(z)$ at $z=0.706$ and $z=0.930$ better than $\Lambda$CDM, simultaneously predicting a larger present-day expansion rate $H_0$, as evident when comparing the red and blue reconstructed curves at $z=0$. \footnote{In the supplementary material, we briefly discuss a comparative study using the DESI and SDSS samples (complete BOSS + eBOSS sample).}

\bigskip
In this \textit{letter}, we have studied the implications of the Baryon Acoustic Oscillations measurements released by the Dark Energy Spectroscopic Instrument for Interacting Dark Energy models characterized by an energy-momentum flow from Dark Matter to Dark Energy. Focusing on the minimal Planck-2018+DESI data combination, we found a preference for interactions exceeding the 95\% confidence level, yielding a present-day expansion rate \textbf{$H_0 = 70.8^{+1.4}_{-1.7}$} km/s/Mpc which can resolve the Hubble tension. Combining Planck-2018+DESI with either measurements of the expansion rate $H(z)$ obtained from the relative ages of massive, early-time, and passively-evolving galaxies or with distance moduli measurements from Type-Ia Supernovae sourced from the Pantheon-plus catalog using the SH0ES Cepheid host distances as a calibrator, we still find a preference for $\xi \neq 0$ at more than 95\% CL. For all the different combinations of datasets, we observe an improvement in the $\chi^2$ of the fit over $\Lambda$CDM. Overall, accounting for DESI data, high and low redshift background measurements are found to be equally or better explained within the IDE framework than $\Lambda$CDM, while consistently yielding values of $H_0$ that are higher than in the standard cosmological model and in much better agreement with local distance ladder estimates. 

In light of these results, we conclude that DESI data (re)open the possibility of addressing the Hubble tension through late-time new physics. However, we emphasize that our analysis of LSS data primarily focused on distance measurements from DESI BAO and Type Ia SN. An important next step is to check whether the model satisfies constraints related to structure formation. In this context, the growth rate of matter density perturbations derived from peculiar velocities, associated with RSD, is commonly used to constrain the combination $f(z)\sigma_8(z)$, which can significantly influence the class of models considered here. Although RSD measurements from the DESI collaboration have not yet been released, we extend our analysis to include 22 measurements of $f\sigma_8(z)$ spanning the redshift range $0.02 < z < 1.944$, gathered from various surveys and summarized in Table I of Ref.~\cite{Sagredo:2018ahx}. The analysis of RSD measurements provides substantial additional information for constraining the IDE framework, thereby strengthening the constraints and limiting the interaction parameter. Using RSD measurements alone, we obtain a constraint of $\xi > -0.033$ (i.e., an order of magnitude better than Planck-2018+DESI), which further improves to $\xi > -0.0133$ when combined with DESI, SN, and CC data (both at the 95\% CL). This latter combination restricts the ability of this class of models to fully resolve the $H_0$ tension, yielding $H_0 = 68.8 \pm 1.3$ km/s/Mpc, and suggests difficulties with structure growth constraints. Although this conclusion is subject to several caveats,\footnote{While robust, some caveats surrounding our analysis of RSD include the reliance on extrapolating linear models into regimes where they may not be valid, the absence of a quasi-linear power spectrum in IDE that could potentially weaken the constraints on $\xi$, and the fact that, while distance measurements come from DESI, the RSD measurements are sourced from various studies in the literature, each employing different approximations and methods.} the analysis of RSD data highlights the need for improved theoretical modeling to address RSD measurements effectively. For a more detailed analysis of the implications of structure formation growth in different IDE regimes (especially concerning the relation between the $H_0$ and $S_8$ tensions), we refer to the companion analysis presented in Ref.~\cite{Sabogal:2024yha}. Therefore, we conclude by stressing that while the IDE model under consideration can resolve the Hubble tension based on background data, it faces challenges with large-scale structure probes sensitive to the dynamics of perturbations and the growth of cosmic structure, highlighting the need for theoretical and phenomenological improvements in this direction.

\begin{acknowledgments}
\bigskip
\noindent The authors express their gratitude to the referee for their valuable comments and suggestions, which have greatly enhanced the overall quality of the work. WG is supported by the Lancaster–Sheffield Consortium for Fundamental Physics under STFC grant: ST/X000621/1. MAS received support from the CAPES scholarship. RCN thanks the financial support from the Conselho Nacional de Desenvolvimento Cient\'{i}fico e Tecnologico (CNPq, National Council for Scientific and Technological Development) under the project No. 304306/2022-3, and the Fundação de Amparo à pesquisa do Estado do RS (FAPERGS, Research Support Foundation of the State of RS) for partial financial support under the project No. 23/2551-0000848-3. EDV is supported by a Royal Society Dorothy Hodgkin Research Fellowship. This article is based upon work from COST Action CA21136 Addressing observational tensions in cosmology with systematics and fundamental physics (CosmoVerse) supported by COST (European Cooperation in Science and Technology). We acknowledge IT Services at The University of Sheffield for the provision of services for High Performance Computing.
\end{acknowledgments}
\bibliographystyle{apsrev4-1}
\bibliography{IDE}
\clearpage
\appendix
\section{Comparing DESI and SDSS constraints}
\label{SDSS}

We present a quantitative comparison between results obtained using DESI and SDSS BAO measurements for the Interacting Dark Energy model studied in this work. In particular, we consider two distinct datasets:
\begin{itemize}

\item \textbf{Planck-2018+DESI:} Planck temperature polarization and lensing likelihoods~\cite{Planck:2019nip,Planck:2018vyg,Planck:2018nkj} in combination with the \textbf{DESI} BAO measurements obtained from observations of galaxies \& quasars~\cite{DESI:2024uvr}, and Lyman-$\alpha$~\cite{DESI:2024lzq} tracers summarized in Tab.I of Ref.\cite{desicollaboration2024desi} (this dataset is labeled as Planck-2018+DESI throughout the main manuscript).

\item \textbf{Planck-2018+SDSS:} Planck temperature polarization and lensing likelihoods~\cite{Planck:2019nip,Planck:2018vyg,Planck:2018nkj} in combination with the \textbf{SDSS} BAO measurements from the final SDSS collaboration compilation encompassing the eight distinct redshift intervals summarized in Table 3 of~\cite{Alam_2021}.
\end{itemize}

We note that both datasets focus on the same distance and expansion rate measurements, namely the isotropic BAO measurements of $D_V(z)/r_{\rm d}$ (where $D_V(z)$ and $r_{\rm d}$ denote the spherically averaged volume distance and sound horizon at baryon drag, respectively), as well as anisotropic BAO measurements of $D_M(z)/r_{\rm d}$ and $D_H(z)/r_{\rm d}$ (with $D_M(z)$ representing the comoving angular diameter distance and $D_H(z) = \frac{c}{H(z)}$ denoting the Hubble distance). Both datasets have similar constraining power; therefore, combining them can be interesting to determine whether there is a consistent preference away from $\Lambda$CDM towards higher $H_0$. Additionally, it will help assess to what extent this preference is driven by new DESI BAO measurements.

\begin{figure}[!h]
\centering
\includegraphics[width=0.8\columnwidth]{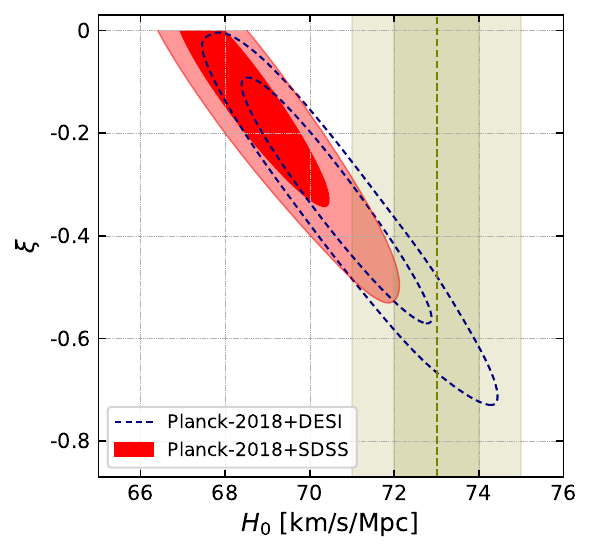}
\caption{2D contours at 68\% and 95\% CL for the coupling parameter $\xi$ and the Hubble parameter $H_0$, as inferred by the different combinations of Planck-2018, DESI, and SDSS data listed in the legend. The olive-green band represents the value of $H_0$ measured by the SH0ES collaboration.}
\label{fig:comparison_sdss_desi}
\end{figure}

Tab.~\ref{tab.results.sdss} summarizes the observational constraints for the main baseline of the model, while Fig.~\ref{fig:comparison_sdss_desi} shows the marginalized probability contours in the $H_0$-$\xi$ plane for both Planck-2018+DESI and Planck-2018+SDSS. From the joint analysis of Planck-2018+SDSS, we find the lower limit $\xi > -0.389$ at 95\% CL, indicating no concrete preference for an interaction in the dark sector. On the brighter side, we still observe a shift towards larger values of $H_0 \sim 69$ km/s/Mpc, which is in better agreement with SH0ES than a minimal $\Lambda$CDM model. In contrast, as discussed in the main text, the analysis of Planck-2018+DESI data yields a substantial shift towards a non-vanishing energy-momentum flow from the DM to the DE sector $\xi = -0.32^{+0.30}_{-0.29}$ at 95\% CL, that further moves the value of $H_0$ towards local distance ladder estimates. Therefore, taking these results at face value, it appears that a significant part of the preference for non-zero $\xi$ comes from DESI BAO data while no preference for $\xi\ne 0$ is not found in SDSS data. However, we would like to emphasize that within the IDE framework of this work, there is no actual tension in the constraints obtained for Planck-2018+SDSS and Planck-2018+DESI. The SH0ES, Planck-2018+DESI, and Planck-2018+SDSS contours all overlap well within the 95\% confidence level in the $H_0$-$\xi$ plane, see also Fig.~\ref{fig:comparison_sdss_desi}.\footnote{In the spirit of discussing all possibilities and cross-referencing our findings with different datasets, we refer readers to~\cite{Bernui:2023byc}, where we tested the same model against some 2D BAO measurements extracted from the SDSS catalog. This dataset has recently gained research attention within the cosmology community. In this case, we observe a preference for a non-zero $\xi$ that aligns well with the results obtained from DESI.}

\begin{table}[!h]
\begin{center}
\renewcommand{\arraystretch}{1}
\resizebox{0.5\textwidth}{!}{
\begin{tabular}{l c c c c c c c c c c c}
\hline
\textbf{Parameter} & \textbf{Planck-2018+DESI } & \textbf{Planck-2018+SDSS}  \\ 
\hline\hline

$ 10^2\times \Omega_\mathrm{b} h^2  $ & $2.243\pm 0.014$ & $2.236\pm 0.013$  \\ 
$ \Omega_\mathrm{c} h^2  $ & $0.079^{+0.025}_{-0.016}$ & $0.101^{+0.016}_{-0.012}$  \\  
$ H_0  $ & $70.8^{+1.4}_{-1.7}$ & $68.92^{+0.96}_{-1.2}$ \\ 
$ \tau_\mathrm{reio} $  & $0.0555\pm 0.0074$ & $0.0544\pm 0.0079$  \\ 
$ \log(10^{10} A_\mathrm{s})$ & $  3.045\pm 0.014$ & $3.045\pm 0.016$  \\ 
$ n_\mathrm{s}  $ & $0.9672\pm 0.0037$ & $0.9650\pm 0.0037$ \\ 
$ \xi  $ & $ -0.32^{+0.18}_{-0.14}\, ( -0.32^{+0.30}_{-0.29} )$ & $ > -0.207 (> -0.389)$ \\ 
\hline \hline
\end{tabular} }
\end{center}
\caption{
\small Constraints at 68\% (95\%) CL on the parameters of the IDE model.}
\label{tab.results.sdss}
\end{table}

\end{document}